\documentclass[10pt,twocolumn,romanappendices]{IEEEtran}

\usepackage[noadjust]{cite}
\usepackage[usenames,dvipsnames]{color}

\usepackage{color}
\usepackage{algorithm,algcompatible,amsmath}
\usepackage{latexsym,amsfonts,amsmath,amssymb,mathrsfs}
\usepackage{verbatim,cite}
\usepackage{graphicx}
\usepackage{floatflt}
\usepackage{subfigure}
\usepackage{color}
\usepackage{setspace}
\usepackage{wrapfig}
\usepackage{empheq}
\usepackage{fonttable}
\usepackage[weather]{ifsym}
\usepackage{stmaryrd}
\usepackage{stackengine}
\usepackage[multiple]{footmisc}
\usepackage[export]{adjustbox}
\def\bth{{\boldsymbol{\theta}}}

\def\bzero{{\boldsymbol{0}}}

\def\c1{{\textcircled{a}}}

\def\bx{{\boldsymbol{x}}}

\def\bC{{\boldsymbol{C}}}

\def\bH{{\boldsymbol{H}}}
\def\bI{{\boldsymbol{I}}}
\def\bJ{{\boldsymbol{J}}}

\def\bN{{\boldsymbol{N}}}
\def\bM{{\boldsymbol{M}}}

\def\bP{{\boldsymbol{P}}}
\def\bQ{{\boldsymbol{Q}}}
\def\bR{{\boldsymbol{R}}}

\def\bU{{\boldsymbol{U}}}
\def\bV{{\boldsymbol{V}}}

\def\bX{{\boldsymbol{X}}}
\def\bY{{\boldsymbol{Y}}}

\def\complexC{{\mathbb{C}}}
\def\realR{{\mathbb{R}}}

\def\bzero{{\boldsymbol{0}}}

\def\@begintheorem#1#2{\tmpitemindent\itemindent\topsep 0pt\rm\trivlist
	\item[\hskip \labelsep{\indent\it #1\ #2:}]\itemindent\tmpitemindent}
\def\@opargbegintheorem#1#2#3{\tmpitemindent\itemindent\topsep 0pt\rm \trivlist
	\item[\hskip\labelsep{\indent\it #1\ #2\
		\rm(#3):}]\itemindent\tmpitemindent}
\def\@endtheorem{\endtrivlist\unskip}

\DeclareMathOperator{\tr}{tr}

\ifCLASSINFOpdf
\else
\fi
\def\bth{{\boldsymbol{\theta}}}

\def\bzero{{\boldsymbol{0}}}

\def\c1{{\textcircled{a}}}

\def\bx{{\boldsymbol{x}}}

\def\bC{{\boldsymbol{C}}}

\def\bH{{\boldsymbol{H}}}
\def\bI{{\boldsymbol{I}}}
\def\bJ{{\boldsymbol{J}}}

\def\bN{{\boldsymbol{N}}}
\def\bM{{\boldsymbol{M}}}

\def\bP{{\boldsymbol{P}}}
\def\bQ{{\boldsymbol{Q}}}
\def\bR{{\boldsymbol{R}}}

\def\bU{{\boldsymbol{U}}}
\def\bV{{\boldsymbol{V}}}

\def\bX{{\boldsymbol{X}}}
\def\bY{{\boldsymbol{Y}}}

\def\complexC{{\mathbb{C}}}
\def\realR{{\mathbb{R}}}

\def\bzero{{\boldsymbol{0}}}

\usepackage{amsmath}
\usepackage{lipsum}

\usepackage{lipsum}
\usepackage{mathtools}

\usepackage{cuted}
\usepackage{amsmath,amssymb,amsthm,mathrsfs,amsfonts,dsfont} 

\algnewcommand\INPUT{\item[\textbf{Input:}]}%
\algnewcommand\OUTPUT{\item[\textbf{Output:}]}%

\newcommand\bss[1]{\boldsymbol{#1}}

\begin{document}
% Do not put or special symbols in the title.

\title{\huge ComSens: Exploiting Pilot Diversity for Pervasive Integration of Communication and Sensing in MIMO-TDD-Frameworks}
\author{Mohammadreza Mousaei, Mojtaba Soltanalian, and Besma~Smida,~\IEEEmembership{Senior Member,~IEEE}\vspace{0.25cm}

Department of Electrical and Computer Engineering, University of Illinois at Chicago (UIC), Chicago, IL, USA\\

Email: \{mmousa3, msol, smida\} @uic.edu \\
}

\maketitle

\begin{abstract}
In this paper, we propose a fully-integrated radar and communication system -- named ComSens. We utilize two different pilot sequences (one for uplink and one for downlink) with the condition that they must be uncorrelated to each other. Within such a framework, the signal received from end-user and the back-scattered signal from the desired objects have uncorrelated pilots. Thus, the base-station is able to distinguish data signal from user and back-scattered signal from object. We assume a time division duplex (TDD) framework. The pilot sequences are designed for MIMO channels. We evaluate channel MSE as a figure of merit for communication system. We also show that the designed pilots are uncorrelated for a range of time lags. Moreover, designed uplink pilot has negligable autocorrelation for a range of time lags leading to an impulse-like autocorrelation for radar sensing.%We employ the pilot overhead -- commonly used in communication systems -- for radar sensing. 
 
\end{abstract}


% \fi
\IEEEpeerreviewmaketitle

\section{Introduction}
Due to the increasing demand in wireless communication
services, achieving higher data rates and more reliable trans-
missions have become a fundamental goal \cite{Alireza1, hajizadeh, Parsa1, MohsenKK1}. Given the ever-increasing demand for both high-speed data services and accurate remote sensing capabilities, modern wireless systems will increasingly require more efficient strategies for use of the available frequency spectrum \cite{Parsa3, Vahid1, Alireza2, Saeed5, Parsa2}. In particular, the coexistence of communication and radar systems has recently attracted a significant research interest \cite{Mohsen1,Vahid3, MohsenKK4, Saeed4, Mohsen2}. For example, different schemes for coexisting communication and radar systems has been proposed; see e.g. \cite{sidelobe, intrapulse, timemod, OFDMRadComm, OFDM2, OFDMRadComExt, RadComMIMO} and the references therein. While integrating radar and communication operation in one system has been considered in the literature, such efforts are typically centered around incorporating communication as a secondary operation alongside a primary radar operation. The research in \cite{sidelobe} exploits the main lobe of the beam for radar purposes, and the sidelobes (which are of no significance to the radar pulse compression) for data transmission purposes. The research works \cite{intrapulse,timemod} approach the same problem by devising similar methods to allow comparably low data rates into an already existing radar system. %Moreover, using OFDM signal structure to construct a joint radar-communication system (RadCom) is discussed in the literature \cite{OFDMRadCom, OFDM2}, mainly by utilizing the distortion of back-scattered OFDM signal compared to transmitted OFDM signal to achieve range and Doppler profiles of the object. OFDM based RadCom is extended for multipath and multiuser scenario in \cite{OFDMRadComExt}. A MIMO OFDM RadCom system is proposed in \cite{RadComMIMO}, in which intelligent OFDM waveform is designed to be suitable for performing both radar sensing and data transmission.  Contrary to former literatures, RadCom has higher data rates but comparably poor radar performance as the OFDM signal is designed for communication purpose.

% The idea of combining communication and radar systems in a single platform has been proposed \cite{idea} and such a system has been developed \cite{implement}. In wireless communication systems, data is transmitted in packets. Each packet has some control bits dedicated to channel estimation (pilot symbols) and rest of the packet contains the data. The goal of this research is to utilize pilot symbols to contribute towards integrating communication and radar systems.

In this work, we  propose an integrated system of communication and sensing (which we call \emph{ComSens}) that relies on  the communication pilot overhead--- thus paving the way for pilot design and exploiting pilot diversity to achieve a satsifactory performance in both communication and radar tasks. Note that:

\emph{Pilot (or training) based channel estimation is very common \cite{me1, hassibi2003much}.} Accurate knowledge of channel state information (CSI) is important for wireless communication systems \cite{Saeed1, MohsenKK2}. Most modern wireless systems acquire the CSI with the assistance of pilot signals (a.k.a. training sequences) that are inserted within the transmit signals periodically \cite{Saeed2, MohsenKK3}. In such scenarios, the transmitter sends training sequences -- known to the receiver -- enabling the receiver to perform channel estimation on the basis of the received training symbols.% \cite{Sayeed, hassibi2003much, biguesh2006training, Giannakis, Liu2007training, bjornson2010training,  shariati2011robust, Nafiseh_TSP, 5551240, tuan2010optimized,noh2014pilot,wang2015channel,xie2015training}.

%Cyber security. The US government set cyber security as a goal -- pursued by the NSF as a strategic research direction [45]-[47]. To this end, it is important not only to use pilot signals serving both radar and communication subsystem, but also to employ signals that are hard to guess by the adversary.

\emph{Communication devices are more ubiquitous than radar systems} \cite{Vahid2, Saeed6}. We note that incorporating the communication signals in the primary radar probing waveforms may not be an efficient fusion of communication and radar systems. In fact, the communication task must play a primary rule not only because of the pervasive usage of comuunication devices, but also the fact that the communication systems typically require a larger capacity of conveying information than radar systems. Additionally, considering the communication operation as the primary lays the ground for making the radar systems ubiquitous (for example having radar capability on cellphones).

%\begin{figure}
%	\centering
%	%\subfigure[Transmit]{	\includegraphics[scale=0.35]{1.pdf}
%	%}
%	%\subfigure[Receive]{	\includegraphics[scale=0.35]{2.pdf}
%	%}
%	\includegraphics[scale=0.75]{ComSens_2.pdf}
%	
%	\caption{The \emph{ComSens} communication and sensing integration framework.}	
%	\label{protocol}
%\end{figure}

\begin{figure*}
	\centering
	%\subfigure[Transmit]{	\includegraphics[scale=0.35]{1.pdf}
	%}
	%\subfigure[Receive]{	\includegraphics[scale=0.35]{2.pdf}
	%}
	\includegraphics[width=18cm]{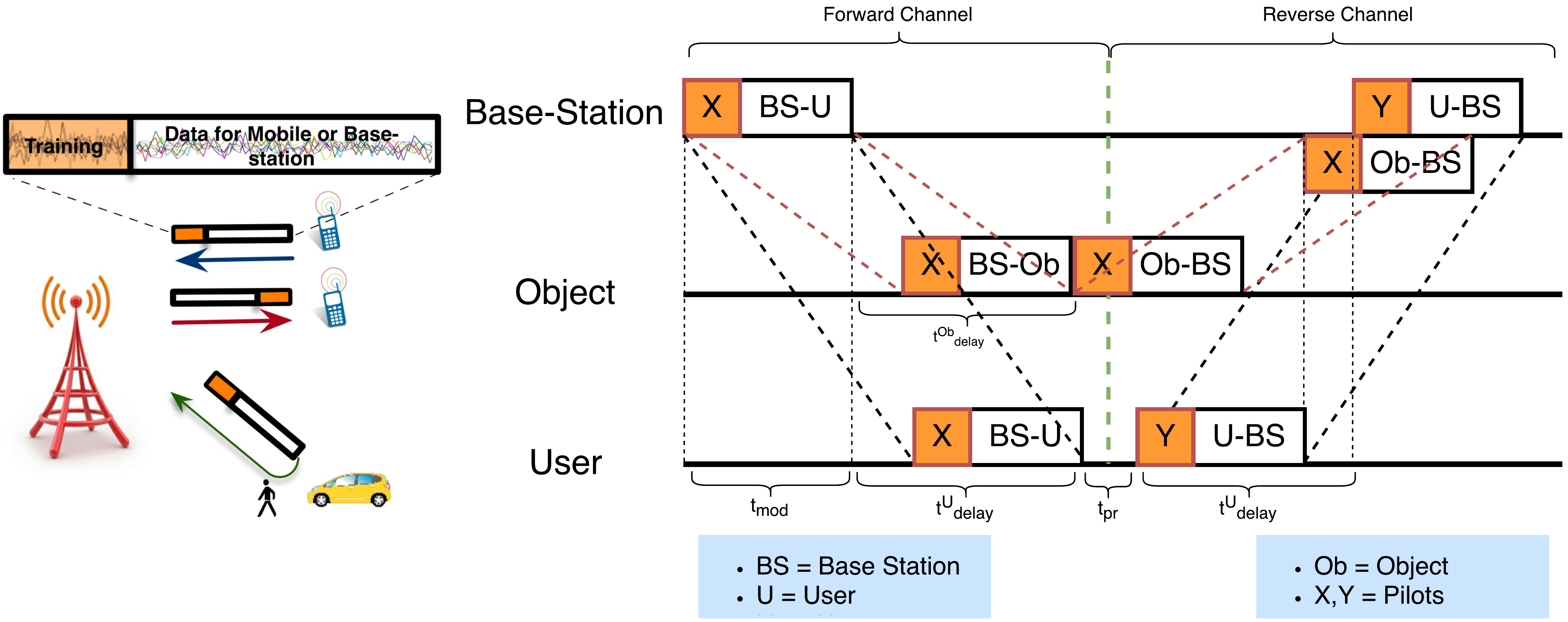}
	
	\caption{The \emph{ComSens} communication and sensing integration framework.}	
	\label{comb}
\end{figure*}

\subsection{Contributions}

The key departure from prior works on integrated radar-communication systems is that we (a) incorporate a radar system in an already existing communication system, particularly by (b) using the novel idea of designing different training signals such that sensing and communication can co-exist. (c) incorporating such a system in TTD framework in a MIMO system and designing the pilot sequences. 

\subsection{Notation}
 We use bold lowercase letters for vectors/sequences and bold uppercase letters for matrices. $(\cdot)^T$, $(\cdot)^*$ and $(\cdot)^H$ denote the vector/matrix transpose, the complex conjugate, and the Hermitian transpose, respectively.  $\| \bx \|_n$ or the $l_n$-norm of the vector $\bx$ is defined as $\left( \sum_k |\bx(k)|^n \right)^\frac{1}{n}$ where  $\{ \bx(k) \}$ are the entries of $\bx$. The Frobenius norm of a matrix $\bX$ (denoted by  $\| \bX \|_F$) with entries $\{ \bX(k,l) \}$ is equal to $\left( \sum_{k,l} |\bX(k,l)|^2 \right)^\frac{1}{2}$.   Finally,  $\realR$ and $ \complexC $ represent the set of real and complex numbers, respectively. %for any real number $x$, the function $[x]$ yields the closest integer to $x$ (the largest is chosen when this integer is not unique). 

%\subsection{Contributions}

%The goal of this paper is to design pilot sequences to contribute towards integrating communication and radar systems.  The key departure of our work is that (1) we use the noble idea of utilizing pilot sequence toward integrating communication and radar systems %inja kar dare
%(2) our proposed system is able to have access to communication and radar systems simultaneously and in the same frequency.

\section{Fusion of Communication and Radar Operations}

In this section, we describe the problem settings and explain our proposed scheme in more details.

\subsection{The Proposed Integration Scheme}
The ComSens framework operates by exploiting the two-way communication between the base-station and end-users.   Before discussing the issue of designing the pilot sequences, we will first address how the base-station and end-user exchange messages and sense the environment at the same time and over the same frequency.  We consider a model with $M$-user multiple-access-broadcast channel (MABC) as depicted in Fig. \ref{comb}. Note that such a system model, in which several end-users wish to exchange messages with a central node, or base-station, is a model that captures the behavior of current and future cellular networks. We assume  half-duplex end-user nodes that may transmit or receive at a given time, on a given frequency, but not both, leading to the need to describe protocols, or which nodes transmit when.  We consider time division duplex (TDD) two-way system as duplex scheme and for multiple access both FDMA and TDMA can be used.  %The transmission time is divided into $M$ time periods, each of which consist of two phases. During the first period, the terminal nodes transmit to the base-station. During the second period the base-station transmits  to the terminal nodes; see Fig. \ref{comb}.
For each user, time is devided into \textit{forward channel} and \textit{reverse channel} (as in TDD scheme). During the former time, Base-station transmits the packet and during the latter user transmits the packet. 
The base-station (BS) sends a packet $s$ to the end-user U.  The end-user can extract its own message after channel estimation using the downlink pilot -- labeled $\boldsymbol{X}$. Contrary to most of the current works on integrated radar and communication systems, the data transmission proposed here is similar to the conventional half-duplex transmission.  This guarantees a high-data rate to efficiently accommodate downlink  traffic. At the same time, the packet $s$ is reflected from objects in the neighborhood. The base-station observes the echo of its own transmit signal, and detects the presence of objects and their distance and relative velocity. With ComSens, the base-station jointly estimates the radar return and extract the uplink message from end-user U after channel estimation using uplink pilot -- labeled $\boldsymbol{Y}$.  The principal constraint in the performance of radar sensing is the simultaneous reception of the radar echo and uplink packet. Therefore the main goal of this work is to design the uplink and downlink pilot sequences.  We design the two pilot sequences  to be uncorrelated to each other so that they can be distinguished from each other at the base-station. After separation of two pilots, the base-station uses the packet with uplink pilot for communication purposes and the reflected downlink packet for sensing.

%{\color{red}The text we had before (to use in modifying the text below): We partition the pilot signals into two parts; each part to be transmitted by a communication agent. Therefore, after an initial pilot signal transmission, the receiver communicates a different signal coming back to the original device of the base station, while a passive object will back-scatter the same signal--- thus paving the way for the system to distinguish various signals, and consequently, being able to perform as a radar system at the same.}

 %As it is shown in Fig.~\ref{protocol}, ComSens consists of a base station, a communication user and an object. First, base station starts by transmitting packets with pilot 1. Mobile user receives the packet with pilot 1, and transmits another packet with pilot 2. The object will back-scatter the same packet with pilot 1. Base station receives a packet with pilot 2 from the user and  a packet with pilot 1 from the object. 

\emph{Remark:} From the above discussion it must be clear that ComSens uplink communications may be subject to interference from the radar echo.  Note that the echo signal is received at the base-station with high attenuation due to the two-way
link  (from the base-station to the object and from the object to the base-station) and the absorption at the object so its impact on the uplink communication is negligible.

\subsection{Time and Range Analysis}

%\begin{figure}
%\centering
%	\includegraphics[width=9cm]{Diagram2.pdf}
%
%	\caption{Time Diagram.}	
%	\label{diagram}
%\end{figure}

\textit{1. Limitations: }

Consider one TDD frame for an end-user U (as it is shown in Fig. \ref{comb}). At the forward channel time, base-station transmits the packet. End-user receives the packet at $t_{mod}+t^U_{delay}$ where $t_{mod}$ is the modulation and transmission time and $t^U{delay}$ is the propagation time between end-user and the base-station. Packet is processed at the end-user in $t_{pr}$ time. Then, end-user transmits the packet in the reverse channel time and base-station receives it at the time $t_1 = t_{mod}+2t^U_{delay}+t_{pr}$. On the other hand, transmitted packet from the base-station is also received at the object at the time $t_{mod} + t^{Ob}_{delay}$ where $t^{Ob}_{delay}$ is the propagation time between object and the base-station. The packet is then back-scattered from the object and received at the base-station at the time $t_2 = t_{mod} + 2t^{Ob}_{delay}$.
We design downlink pilot  and uplink pilot to be uncorrelated to each other for $k$ time lags. Therefore, if two received signals (from user and object) have arrival time difference ($t_2-t_1$) of at most $k$, they are distinguishable from each other. On the other hand, if $t_2-t_1 > k$, the radar signal cannot be recognized and it will be considered as weak interference for communication system. Consequently, our proposed integrated radar system will perform when $t_2-t_1 \leq k$. Substituting $t_1$ and $t_2$ we have:

\begin{eqnarray}
\label{RadCond}
t^{U}_{delay} - t^{Ob}_{delay} \leq \dfrac{t_{pr}+k}{2}
\end{eqnarray}
where
\begin{equation}
\label{del1}
t^{Ob,U}_{delay} = \dfrac{d^{Ob,U}}{\nu T_s}
\end{equation}
%\begin{equation}
%\label{del2}
%t_{delay2} = \dfrac{d_{2}}{\nu T}
%\end{equation}
and $d^{U}$ and $d^{Ob}$ are respectively the distance of user and the object from the base-station, $T_s$ is symbol time in our system and $\nu$ is the speed of electromagnetic wave in the space. Using Eq. (\ref{RadCond}, \ref{del1}) we have

\begin{equation}
\label{notrange}
 d^{Ob} \leq  d^{U} + \dfrac{\nu T (t_{pr} + k)}{2} 
\end{equation}
%if $d_2$ is less than the lower bound 

%Assume that we have $N$ symbols to transmit (since we have a MIMO system, here $N = N_s/n_t$ where $N_s$ is number of symbols). Transmission schedule for our system is shown in Fig. \ref{comb}. Packet sent from base-station to end-user 1 is received at the time $N + t_{delay1}$ (the process of transmission takes $N$ blocks of time and we define $t_{delay1}$ as the time distance between base-station and end-user 1). This transmitted packet is also received at the object at the time $N + t_{delay2}$ ($t_{delay2}$ is defined as the time distance between base-station and object). Now packet received at end-user 1 is processed (we assume a processing time of $t_p$) and then a packet is transmitted back to base-station. This packet is received at base-station at the time $N + 2t_{delay1} + t_p$, while the back-scattered packet (from the object) is received at the base-station at the time $N + 2t_{delay2}$. Then, the same scenario happens to end-user 2.  

\textit{2. Practical Scenario: } Communication cell towers have a range between $35km$ to $72km$. We consider our user to be (as a medium distance) at the distance $d^{U} = 25km$ of the base-station. %As it is shown in Fig. \ref{range}, for a fixed user, we have a blind side for radar (white region) and a working radar region (green region).
 Assume that the symbol time $T_s = 25 \mu s$  and processing time $t_{pr} = T_s$ where speed of electromagnetic wave is $\nu = 3\times10^{8}$, assuming we design our pilots to be uncorrelated for $k = 4$. Such a system would have a radar range of 43.75km ($d^{Ob} \leq 43.75km$).

\subsection{Channel Model}
We consider the same settings as in \cite{Nafiseh_TSP} and \cite{soltan}. More precisely, we consider a narrowband block fading point-to-point  MIMO channel with $n_T$ transmit and $n_R$ receive antennas. Assume that $\bP \in \mathds{C}^{B\times n_T}$ be a matrix whose rows are the pilot sequence at each transmitter antenna. At the training phase, channel can be described as

\begin{equation}
\bY = \bH\bP^T + \bN
\end{equation}  
where $\bY \in \mathds{C}^{n_R\times B}$ is the received sequence, $\bH \in \mathds{C}^{n_R\times n_T}$ is the MIMO channel when $\bH(i,j)$ denotes the MIMO channel gain between $i^{th}$ transmitter and $j^{th}$ receiver and $\bN \in \mathds{C}^{n_R\times B}$ is the noise matrix. We assume Gaussian noise i.e. $vec(\bN) \sim \mathcal{CN}(\textbf{0}, \bM) $ where $\bM \in \mathds{C}^{Bn_R\times Bn_R}$ denotes noise covariance matrix. We also assume $vec(\bH) \sim \mathcal{CN}(\textbf{0}, \bR)$ where $\bR \in \mathds{C}^{n_Tn_R\times n_Tn_R}$ denotes channel covariance matrix.

\begin{figure}
\centering
	\includegraphics[width=9cm]{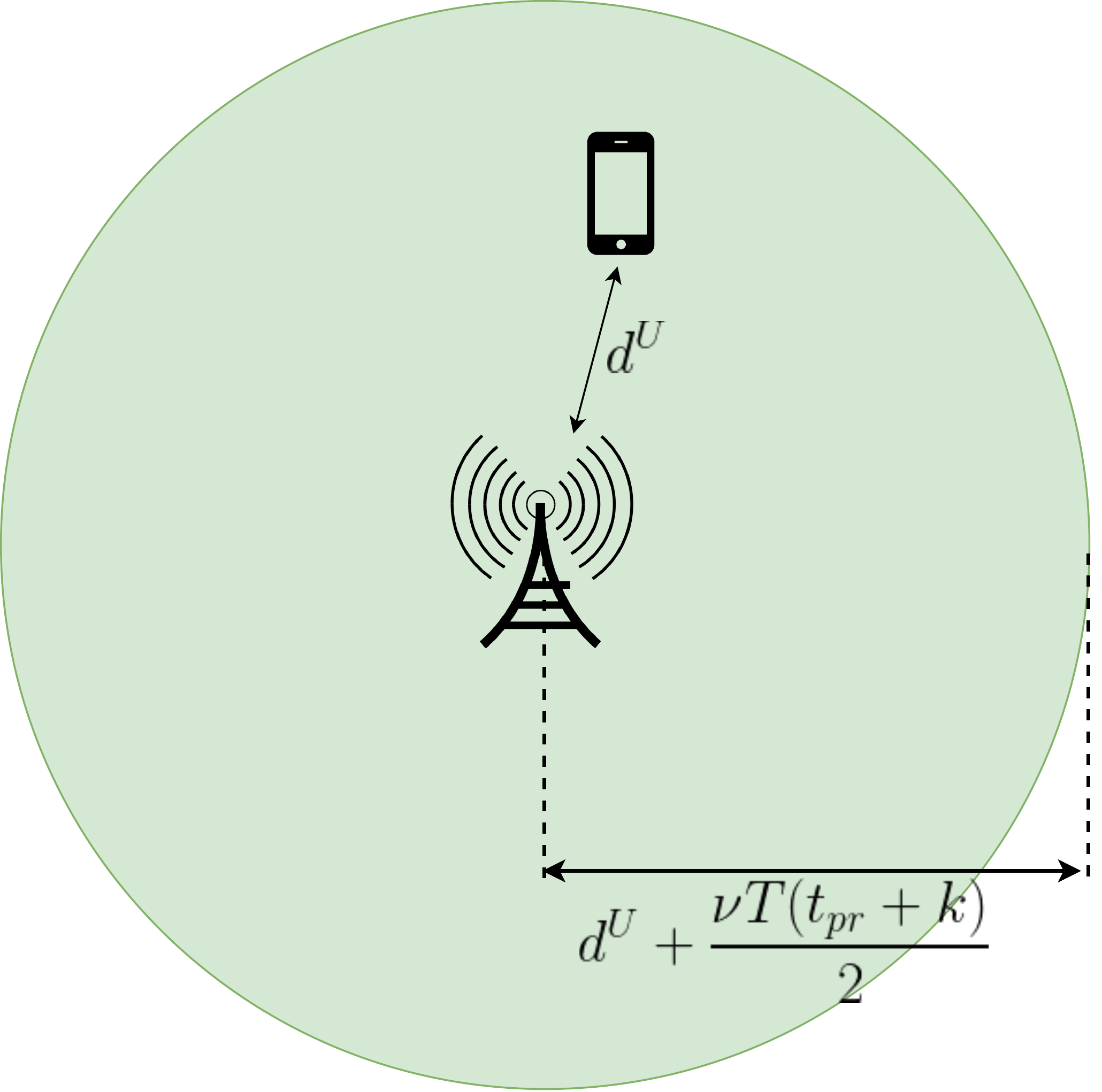}

	\caption{The radar operational range of ComSens.}	
	\label{range}
\end{figure}

\section{Pilot Sequence Design}

In this section, we design the pilot coefficients gathered in the matrix $\bss{P}$, in order to produce an accurate estimate of the channel $\bss{H}$--while simultaneously satisfying a set of radar performance criteria. For an accurate channel estimation, one may resort to a  minimization of the channel mean-squared error (MSE), expressed as \cite{Nafiseh_TSP,soltan}
\begin{eqnarray} \label{eq:MSE}
MSE=%tr\left[\theta\right] \nonumber\\ 
tr\left[ \left(\boldsymbol{R}^{-1} + (\boldsymbol{P}\otimes \boldsymbol{I}_{n_R})^H \boldsymbol{M}^{-1}(\boldsymbol{P}\otimes \boldsymbol{I}_{n_R}) \right)^{-1} \right].
\end{eqnarray}
Let $\widetilde{\bP}  \triangleq {\bf P} \otimes {\bf I}_{n_{R}} \in \complexC^{B n_R  \times n_T n_R}$, and note that using the matrix inversion lemma we have 
\begin{eqnarray}
\bth &\triangleq& \left( \bR^{-1} + \widetilde{\bP}^H \bM^{-1} \widetilde{\bP} \right)^{-1} \\  &=& \bR - \bR \widetilde{\bP}^H \left( \bM + \widetilde{\bP} \bR \widetilde{\bP}^H \right)^{-1} \widetilde{\bP} \bR ,  % \nonumber
\end{eqnarray}
where $MSE = tr[\theta]$. Now let
\begin{equation}\label{eq:def1}
\bQ \triangleq \left(
\begin{array}{cc}
\bR & \bR  \widetilde{\bP}^H \\%\nonumber
\widetilde{\bP} \bR & \bM + \widetilde{\bP} \bR \widetilde{\bP}^H
\end{array}\right) \in \complexC^{(B+n_T)n_R  \times (B+n_T)n_R},
\end{equation}
\begin{eqnarray}
\bU \triangleq (\bI_{n_T n_R}\;\; \bzero_{n_T n_R \times B n_R})^T \in \complexC^{(B+n_T)n_R  \times n_T n_R},
\end{eqnarray}
and observe that \cite{petersen2008matrix},
\begin{equation}\label{eq:def2}
\bU^H \bQ^{-1} \bU=\bth^{-1}.
\end{equation}
In light of the above, the authors in \cite{soltan} propose a cyclic optimization approach to minimizing the MSE in \eqref{eq:MSE}: Consider an auxiliary variable $\boldsymbol{V} \in \mathds{C}^{n_Tn_R \times Bn_R}$ such that
\begin{equation}
\label{fvx}
F(\boldsymbol{V}, \boldsymbol{P}) := tr\left[\boldsymbol{V}^H\boldsymbol{Q}\boldsymbol{V}\right].
\end{equation}
 The minimizer $\boldsymbol{V}$ of (\ref{fvx}) can be obtained as \cite[p. 354]{peterbook-spectral}

\begin{eqnarray}
\label{minimizer}
\boldsymbol{V}_* =
\begin{pmatrix}
\boldsymbol{I}_{n_Tn_R} \\
-\left(\boldsymbol{M} + \widetilde{\boldsymbol{P}}\boldsymbol{R}\widetilde{\boldsymbol{P}}^H \right)^{-1}\widetilde{\boldsymbol{P}}\boldsymbol{R}
\end{pmatrix}
\end{eqnarray}
By substituting (\ref{minimizer}) in (\ref{fvx}), one can verify that
\begin{equation}
F(\boldsymbol{V}_*, \boldsymbol{P}) = tr\left[\bth\right] = MSE.
\end{equation}
Therefore, in order to optimize the MSE we can use a cyclic optimization of (\ref{fvx}) with respect to $\bV$ and $\bP$. In particular, it was shown in \cite{soltan} that the optimization of  (\ref{fvx}) with respect to  $\bP$ can be cast at each (cyclic) iteration as:

\vspace{0.03cm}
\begin{equation}
\label{Piteration}
\min_{\boldsymbol{P}^{h+1} \in \Omega} \left|\left| \boldsymbol{P}^{(h+1)} - \boldsymbol{P}_{\Sigma}^{(h)} \right|\right|^2_{2},
\end{equation} 
\vspace{0.03cm}

\noindent where $\boldsymbol{P}_{\Sigma}^{(h)}$  is constructed from $\boldsymbol{P}^{(h)}$ at each iteration (see \cite{soltan} for details). For the two-part pilot employed in ComSens,  define:
\begin{eqnarray}
\boldsymbol{P}_{DL} &:=& \boldsymbol{X} \\
\bP_{UL} &:=& \boldsymbol{Y}
\end{eqnarray}
\vspace{0.03cm}

\noindent where $\boldsymbol{X} \in \mathds{C}^{B\times n_T}$ is the downlink pilot  contributing at both radar and communication modes and $\boldsymbol{Y} \in \mathds{C}^{B\times n_R}$ is the uplink pilot which contributes only in communication mode. Thus, (\ref{Piteration}) becomes

\vspace{0.03cm}
\begin{eqnarray}
\label{optimiz}
\min_{\bX, \bY \in \Omega} \left|\left| \bX - \bX_{\Sigma}\right|\right|_2 ^2+ \left|\left| \bY - \bY_{\Sigma}\right|\right|_2^2,
\end{eqnarray}
\vspace{0.03cm}

\noindent where the constraint set $\Omega$ is yet to be defined. As indicated earlier, $\bX$ and $\bY$ should have \emph{low correlation} with each other and $\bX$ should have an \textit{impulse-like} autocorrelation. We describe the pilot constraints in three categories:
%\vspace{-1cm} 
\begin{enumerate}
\item Both pilot sequences should have fixed transmit powers given by
\begin{eqnarray}
 || \bss{x}_q ||_2^2 \leq p, \quad 1 \leq q \leq n_T\\
 || \bss{y}_l ||_2^2 \leq p, \quad 1 \leq l \leq n_R
\end{eqnarray}
where $\bss{x}_q$ and $\bss{y}_l$ are column vectors of $\bX$ and $\bY$ and $p$ is the power upper-bound.
\item To resolve ambiguity between radar reflections and communication signals, pilot sequences (and their time lags up to $k$ lags) should be uncorrelated to each other; i.e their cross correlation must be zero or very small at least for a number of time lags (forming a zero correlation zone \cite{fan1999class, torii2004new}): \\
\begin{equation}
 \bss{X}^T \bss{J}_{i} \bss{Y} \simeq \bzero^{n_T\times n_R}, \quad 0 \leq i \leq k,
\end{equation}
where $\bss{J}_{k} \in C^{B \times B}$ is a shift matrix that shifts a matrix by $k$ time lags. Clearly $\bss{J}_0$ is identity matrix.
\item Radar pilot sequence should be impulse-like; i.e. its auto-correlation must be zero or very small at least for a number of time lags:\\
\begin{equation}
 \bss{X}^T \bss{J}_{i}\bss{X} \simeq \bzero^{n_T\times n_T},  \quad 0 \leq i \leq k.
\end{equation}
\end{enumerate}  

Consequently, one can solve the following optimization problem to design our pilot sequences:

\begin{eqnarray}
\label{OP}
\min_{\bX, \bY \in \Omega}&& \left|\left| \bX - \bX_{\Sigma}\right|\right|_2 ^2+ \left|\left| \bY - \bY_{\Sigma}\right|\right|_2^2 \\
s.t. \quad  &&|| \bss{x}_q ||_2^2  \leq p, \quad 1 \leq q \leq n_T; \nonumber\\
 &&|| \bss{y}_l ||_2^2 \leq p,\quad 1 \leq l \leq n_R; \nonumber\\
 && \bss{x}_q^T \bss{J}_{i}\bss{y}_l\leq \epsilon,  \quad 1 \leq i \leq k; \nonumber\\
  &&\bss{x}_q^T \bss{J}_{i}\bss{x}_q\leq \epsilon,  \quad 1 \leq i \leq k; \nonumber
\end{eqnarray}
where $\epsilon$ is a very small number (in this paper we use $10^{-5}$) to achieve equality constraints. In order to tackle (\ref{OP}) we can use cyclic optimization\cite{stocia2004cyclic}. We define:

%In order to solve optimization problem Eq. (\ref{Convex}), we first make it an optimization problem with vector constraints instead of matrix constraints.
\begin{equation}
G(\bX,\bY) := \left|\left| \bX - \bX_{\Sigma}\right|\right|_2 ^2+ \left|\left| \bY - \bY_{\Sigma}\right|\right|_2^2
\end{equation} 
Then one can perform a cyclic procedure to minimize $G(\bX,\bY)$ as follows: We start with an initial value $\bY = \bY^0$. Then we comupte $\bX^i$ by tackling minimization problem in Eq. (\ref{XConv}) and $\bY^i$ by tackling minimization problem in Eq. (\ref{YConv}). %until a stop criterion ($||\bX^{i}-\bX^{i-1}||_2+||\bY^{i}-\bY^{i-1}||_2 < \zeta$ for some $\zeta > 0$) is satisfied. 
More precisely:

%sdadas sdasd asda sdadas sdasd asda sdadas sdasd asda sdadas sdasd asda sdadas sdasd asda sdadas sdasd asda sdadas sdasd asda sdadas sdasd asda
\begin{eqnarray}
\label{XConv}
\bX^i &=& arg \min_{\bX} G(\bX,\bY^{i-1}) \\
s.t. &&  || \bss{x}_q ||_2^2  \leq  p, \quad\quad\quad 1 \leq q \leq n_T; \nonumber\\
 && \bss{x}_q^T \bss{J}_{m}\bss{y}^{i-1}_l =  0,  \quad 1 \leq m \leq k; \nonumber\\
  &&\bss{x}_q^T \bss{J}_{m}\bss{x}_q\leq \epsilon,  \quad\quad 1 \leq m \leq k; \nonumber
\end{eqnarray}
\begin{eqnarray}
\label{YConv}
\bY^i &=& arg \min_{\bY} G(\bX^i,\bY) \\
s.t. &&|| \bss{y}_l ||_2^2 \leq p, \quad\quad\quad  1 \leq l \leq n_T; \nonumber\\
&& (\bss{x}^i_q)^T \bss{J}_{m}\bss{y}_l\leq 0,  \quad 1 \leq m \leq k; \nonumber
\end{eqnarray}
%Moreover, constraint 3 is changed to make a convex constraint. Eq. (\ref{OP}) is now a convex optimization problem with convex constraints which can be solved using convex optimization techniques.
where $1 \leq q \leq n_T$ and $1 \leq l \leq n_R$. Note that since now the second constraint in both (\ref{XConv}) and (\ref{YConv}) are affine constraints, we replaced them with equality. Eq. (\ref{YConv}) is now a convex optimization problem and solvable using convex optimization. However, the third constraint in (\ref{XConv}) is not convex. We can rewrite Eq. (\ref{XConv}) in form:

\begin{eqnarray}
\label{XConv2}
\bX^i &=& arg \min_{\bX} G(\bX,\bY^{i-1}) \\
s.t. &&  || \bss{x}_q ||_2^2  \leq  p, \quad\quad\quad\quad\quad\quad\quad\quad\quad\quad  1 \leq q \leq n_T; \nonumber\\
 && \bss{x}_q^T \bss{J}_{m}\bss{y}^{i-1}_l =  0,  \quad\quad\quad\quad\quad\quad\quad\quad 1 \leq m \leq k; \nonumber\\
  &&\bss{x}_q^T (\bss{J}^T_{m}+\bss{J}_{m}+2\bss{I}_{m})\bss{x}_q\leq 2p,  \quad\quad 1 \leq m \leq k; \nonumber
\end{eqnarray}
the third constraint in (\ref{XConv2}) is now in quadratic convex form since $(\bss{J}^T_{m}+\bss{J}_{m}+2\bss{I}_{m})$ is a symmetric positive semi-definite matrix. Note that the optimization problem is still the same (since $\bss{x}_q^T \bss{J}_{m}\bss{x}_q\leq \epsilon$ and $\bss{J}^T_{m} = \bss{J}_{-m}$ then $\bss{x}_q^T \bss{J}^T_{m}\bss{x}_q\leq \epsilon$ also holds and from the first constraint $\bss{x}_q^T \bss{I}_{m}\bss{x}_q\leq p$). Now we can follow the steps of the algorithm below to design the pilot sequence.
\begin{algorithm}
\label{algo}
    \caption{Cyclic Algorithm For Constrained Pilot Sequence Design}
    \textbf{Step 0}: Initialize $\boldsymbol{P}_{DL}$ and $\bP_{UL}$ using a random matrix in $\Omega$.\\
    \textbf{Step 1}: Compute the minimizer $\boldsymbol{V}$ of (\ref{fvx}) using (\ref{minimizer}). \\ %using \ref{} 
    \textbf{Step 2}: Update the current design of $\bX$ and $\bY$ by solving cyclic optimization problem (\ref{XConv2}) and (\ref{YConv}) $\mu$ times (or until convergence).\\
    \textbf{Step 3}: Repeat steps 1 and 2 until a stop criterion is satisfied, e.g. $\left|MSE^{(m + 1)} - MSE^{(m)}\right|<\eta$ for some given $\eta > 0$, where $m$ denotes the outer loop iteration. 
\end{algorithm}

\section{Simulation Results}

In this section, we evaluate the performance of the communication method with respect to the channel MSE metric and for the radar mode we illustrate the cross and auto-correlation between two pilot sequences.

\subsection{Simulation Settings}

We used the exponential model to generate covariance matrices. This model is particularly appropriate whenever a control over correlation is required. For a covariance matrix $\bC$, we let $[\bC]_{k,l} = \rho^{l-k}$ for $k \leq l$, and $[\bC]_{k,l} = [\bC]_{l,k}^*$ for $l < k$, with $|\rho| < 1$ denoting the correlation coefficient. Furthermore, we assume that both the channel matrix $\mathbf{R}$ and the noise matrix $\mathbf{M}$ follow the Kronecker model; i.e for covariance matrix $\mathbf{R}$ defined as $\mathbf{R} = (\mathbf{R}^T_{\mathbf{T}} \otimes \mathbf{R_R})$ we suppose $\rho_{rt} = 0.9 e^{- j \theta_{rt} }$ and $\rho_{rr} = 0.65 e^{- j \theta_{rr}}$ to construct $\mathbf{R_T}$ and $\mathbf{R_R}$ (at the transmit side and the receive side, respectively) using exponential model. Also, for covariance matrix of noise $\mathbf{M}$ defined as $\mathbf{M} = (\mathbf{M}^T_{\mathbf{T}} \otimes \mathbf{M_R})$ where $\mathbf{M_R} = \mathbf{R_R}$, we let $\rho_{mt} = 0.8 e^{- j \theta_{mt}}$ to construct $\mathbf{M_T}$ at the transmitter side.The phase arguments $(\theta_{rt},\theta_{rr},\theta_{mt} )$ appearing above were chosen randomly as $(0.8349\pi, 0.4289 \pi, 0.5361 \pi)$. 

We also normalize $\mathbf{R}$ and $\mathbf{M}$ such that $\tr\{\mathbf{R}\} = 1$ and $\tr\{\mathbf{M}\} = 1$, and define the pilot sequence-to-noise ratio (SNR) as 
$
\textrm{SNR} \triangleq \gamma
$, 
 and $\gamma=\| \bP \|_F^2$ denotes the \textit{total training energy}. 
 %Then for a given $\gamma$, different values of $\kappa$ realize different SNR values. 
 We consider $\gamma =  B n_T$, and set the stop threshold of the iteration loop in Algorithm I as  $\eta = 10^{-5}$.

\subsection{Channel MSE Metric} %}{Channel Estimate MSE and Design Criteria}

\begin{figure}
	\begin{center}
		\includegraphics[width=9.5cm]{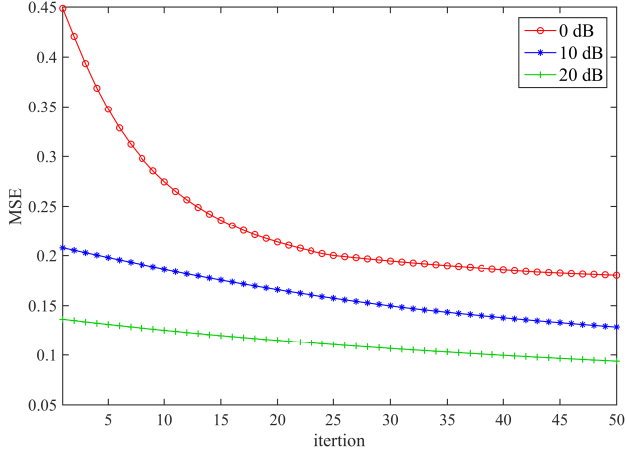}
		\caption{MSE comparison of different schemes for a $4 \times 4$ MIMO channel where $B=8$.  In order to demonstrate the improvement of MSE values through the iterations, values of MSE  at each iterations is shown.} \label{fig:Fig1}
	\end{center}
\end{figure}

We show the performance of the suggested approach for communication purposes using MSE as the figure of merit.  We consider a $4 \times 4$ MIMO channel with $B=8$. The results are shown in Fig.~\ref{fig:Fig1}. For each power, we have used the proposed method $ 50 $ times ,using different initializations, and have reported the average of the obtained MSE values. It can be observed from Fig.~\ref{fig:Fig1} that the proposed method performs better in each iteration until it converges to the optimal MSE.

\subsection{Radar Pilot Sequence Specifications}

\begin{figure}
	\begin{center}
		\includegraphics[width=9cm]{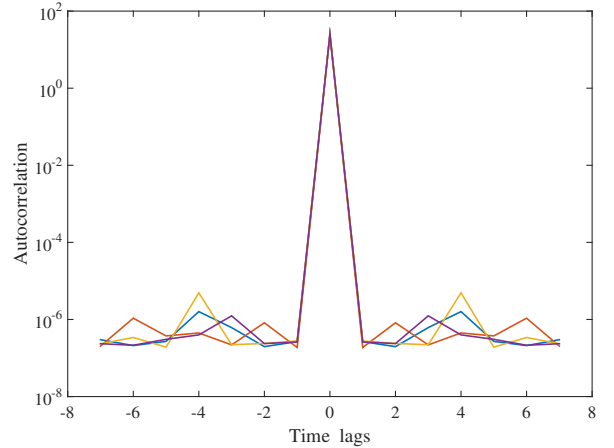}
		\caption{Autocorrelation of radar pilot signal ($\bss{x}_l^T\bJ_i\bss{x}_l$) where $1\leq l\leq 8$, and each $l$ denotes a transmit antenna so we have totally 4 autocorrelation plots in this figure, and also  $-8\leq i\leq 8$ denote time lags} \label{fig:Fig2}
	\end{center}
\end{figure}

To ensure radar part of the system performs properly, our pilot sequence for radar part should have very small auto-correlation for at least a range of time lags so that this pilot sequence have an impulse like shape. In Fig. \ref{fig:Fig2} this auto-correlation is shown. For each lag, auto-correlation level is shown in dB. Fig. \ref{fig:Fig2} shows that auto-correlation levels for time lags 2-8 are almost zero compared to autocorrelation for the first lag. Which gives us the impulse-like correlation for the pilot sequence contributing in sensing mode.

\subsection{Correlation of pilots}

\begin{figure}
	\begin{center}
		\includegraphics[width=9cm]{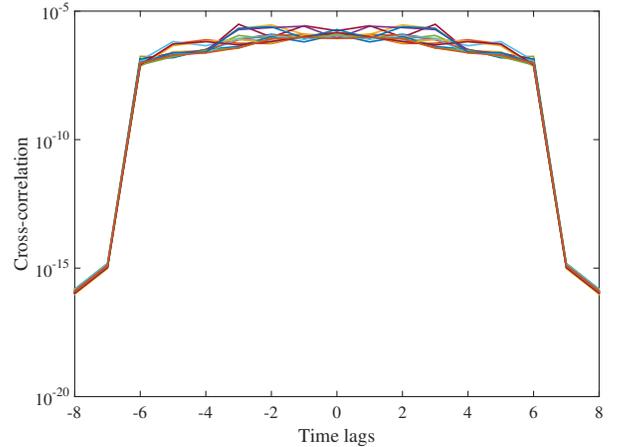}
		\caption{Cross-correlation of radar and communication pilot signals ($\bss{x}_q^T\bJ_i\bss{y}_l$) where $1\leq q\leq 8$, and each $q$ denotes different transmit antenna, $1\leq l\leq 8$, and each $l$ denotes different receive antenna so we have totally 16 cross-correlation plots in this figure, and also  $-8\leq i\leq 8$ denote time lags} \label{fig:Fig3}
	\end{center}
\end{figure}

The key factor for our system to distinguish between radar signal and communication signal is that two pilot sequences should be uncorrelated with each other for a number of time lags. Fig. \ref{fig:Fig3} shows cross-correlation between two pilot signals for our simulations in dB. As it is obvious from simulations results, correlation between these two signals are really small so they can be assumed uncorrelated.

\section{Conclusion}
The idea of designing pilot sequences for a communication system to be able to operate also in an integrated radar mode has been proposed and the protocol and limitations has been explained. We evaluate the channel MSE for communication system and show that pilot sequences are uncorrelated, and also one of the sequences have impulse-like correlation (suitable for radar sensing). The proposed system can perform as a radar and communication system. Considering the communication operation as the primary also
lays the ground for making the radar systems ubiquitous.


%\section*{Acknowledgment}




% can use a bibliography generated by BibTex as a .bbl file
% BibTex documentation can be easily obtained at:
% The IEEEtran BibTex style support page is at:
%\bibliographystyle{IEEEtran}
% argument is your BibTex string definitions and bibliography database(s)
%\begin{thebibliography}{1}
%\end{thebibliography}
\bibliographystyle{IEEEtran}
\bibliography{refs_radar,refs_MS,MIMO_training}{}

\end{document}